\documentclass[11pt,a4paper]{article}

\usepackage[utf8]{inputenc}
\usepackage[T1]{fontenc}
\usepackage{geometry}
\usepackage{amsmath,amssymb,amsfonts}
\usepackage{graphicx}
\usepackage{booktabs} % For professional tables
\usepackage{xcolor}
\usepackage{hyperref}
\usepackage{authblk}
\usepackage{float}
\usepackage{caption}
\usepackage{subcaption}
\usepackage{multirow}
\usepackage{cite}     % Better citation handling
\usepackage{tikz}     % For drawing the Deployment Architecture
\usetikzlibrary{shapes.geometric, arrows, positioning, fit, calc, shadows}

\hypersetup{
    colorlinks=true,
    linkcolor=blue,
    filecolor=magenta,      
    urlcolor=cyan,
    citecolor=red,
    pdftitle={TAG-HGT: Inductive Cold-Start Recommendation},
    pdfauthor={Zhexiang Li}
}

\usepackage{filecontents}

\title{\textbf{TAG-HGT: A Scalable and Cost-Effective Framework for Inductive Cold-Start Academic Recommendation}}

\author[1]{Zhexiang Li}
\affil[1]{Tongji University, Shanghai, China}
\affil[1]{\texttt{lzx1011@tongji.edu.cn}}

\date{\today}

\begin{document}

\maketitle

% ================= Abstract =================
\begin{abstract}
\textbf{Inductive cold-start recommendation remains the ``Achilles' Heel'' of industrial academic platforms}, where thousands of new scholars join daily without historical interaction records. While recent Generative Graph Models (e.g., HiGPT \cite{tang2024higpt}, OFA \cite{liu2024ofa}) demonstrate promising semantic capabilities, their prohibitive inference latency (often exceeding 13 minutes per 1,000 requests) and massive computational costs render them \textbf{practically undeployable} for real-time, million-scale applications. 

To bridge this gap between generative quality and industrial scalability, we propose \textbf{TAG-HGT}, a cost-effective neuro-symbolic framework. Adopting a decoupled ``Semantics-First, Structure-Refined'' paradigm, TAG-HGT utilizes a frozen Large Language Model (DeepSeek-V3 \cite{deepseek2024v3}) as an offline semantic factory and distills its knowledge into a lightweight Heterogeneous Graph Transformer (HGT \cite{hu2020heterogeneous}) via Cross-View Contrastive Learning (CVCL). 

We present a key insight: while LLM semantics provide necessary \textit{global recall}, structural signals offer the critical \textit{local discrimination} needed to distinguish valid collaborators from semantically similar but socially unreachable strangers in dense embedding spaces. Validated under a strict \textbf{Time-Machine Protocol} on the massive OpenAlex dataset \cite{priem2022openalex}, TAG-HGT achieves a SOTA System Recall@10 of \textbf{91.97\%}, outperforming structure-only baselines by 20.7\%. 

Most significantly, from an industrial perspective, TAG-HGT reduces inference latency by five orders of magnitude ($\mathbf{4.5 \times 10^{5}\times}$) compared to generative baselines (from 780s down to \textbf{1.73 ms}), and slashes inference costs from $\sim$\$1.50 to \textbf{$<$ \$0.001} per 1k queries. This 99.9\% cost reduction democratizes high-precision academic recommendation.

\end{abstract}

% ================= 1. Introduction =================
\section{Introduction}
\label{sec:intro}

Interdisciplinary research has become the engine of modern scientific discovery, exemplified by the ``AI for Science'' paradigm. In large-scale academic platforms (e.g., Google Scholar, ResearchGate), accurately recommending collaborators is critical. However, these systems face a severe \textbf{Inductive Cold-Start Problem}: thousands of new scholars join daily without historical interaction records. Traditional structure-based Graph Neural Networks (GNNs) such as GCN \cite{kipf2016semi} and GAT \cite{velickovic2018graph} heavily rely on existing topological connectivity. Even specialized heterogeneous models like HAN \cite{wang2019heterogeneous} and HGT \cite{hu2020heterogeneous} suffer from the "topological void," often collapsing into random guessing when applied to isolated nodes.

A fundamental challenge is the \textbf{"Dense Semantic Space" dilemma}. While LLMs (e.g., GPT-3 \cite{brown2020language}, Llama 2 \cite{touvron2023llama}) can retrieve hundreds of candidates with similar research interests (Global Recall), they struggle with \textbf{Local Discrimination}. In specialized fields, potential collaborators and random strangers often share nearly identical semantic embeddings. Structure, therefore, acts as the necessary scalpel to separate viable collaborators from semantically similar strangers.

Recent Generative Graph Models (e.g., P5 \cite{geng2022recommendation}, HiGPT \cite{tang2024higpt}) introduce catastrophic computational bottlenecks. As demonstrated in our industrial stress test, the inference latency of autoregressive models often exceeds 780 seconds per 1,000 users. To bridge this gap, we propose \textbf{TAG-HGT}, which reimagines the pipeline as Implicit Knowledge Distillation, leveraging the efficiency of contrastive frameworks \cite{chen2020simple} to transfer the wisdom of DeepSeek-V3 \cite{deepseek2024v3} into a lightweight GNN.

% ================= 2. Related Work =================
\section{Related Work}
\label{sec:related}

\subsection{Evolution of Graph Neural Networks}
Heterogeneous Information Networks (HINs) require specialized encoders \cite{shi2016survey, wu2020comprehensive}. Early works like metapath2vec \cite{dong2017metapath2vec} utilized random walks, while HAN \cite{wang2019heterogeneous} introduced hierarchical attention. HGT \cite{hu2020heterogeneous} generalized this with node-type dependent attention. Fundamental architectures such as GCN \cite{kipf2016semi}, GAT \cite{velickovic2018graph}, and GraphSAGE \cite{hamilton2017inductive} laid the groundwork, while node2vec \cite{grover2016node2vec} provided scalable feature learning. Recently, SeHGNN \cite{yang2023sehgnn} and HGNN \cite{zhang2019heterogeneous} optimized these for efficiency. However, these topology-dependent models struggle in inductive settings where new nodes lack edges.

\subsection{Industrial-Scale Recommender Systems}
Scalability is paramount in industry. Deep learning revolutionized this field with YouTube DNN \cite{covington2016deep} and Wide \& Deep \cite{cheng2016wide}. Graph-based industrial systems followed, including PinSage \cite{ying2018graph} at Pinterest and AliGraph \cite{zhu2019aligraph} at Alibaba, which optimized sampling for billion-scale graphs. LightGCN \cite{he2020lightgcn} and NGCF \cite{wang2019neural} simplified GCNs by removing non-linearities. Our work aligns with this philosophy, utilizing PyTorch Geometric \cite{fey2019fast} to implement a lightweight student model for online serving.

\subsection{Generative Recommendation \& Semantics}
Integrating LLMs into RecSys is a growing trend \cite{lin2023llmrecsys, wu2023survey}. Models like TAPE \cite{he2023explanations} use LLMs to enhance node features. Recent works like HiGPT \cite{tang2024higpt}, OFA \cite{liu2024ofa}, and TALLM \cite{bao2023tallm} attempt to fine-tune LLMs on graph instructions. Our approach leverages pre-trained representations—ranging from early Word2Vec \cite{mikolov2013distributed} and GloVe \cite{pennington2014glove}, to BERT \cite{devlin2018bert, reimers2019sentence}, and finally to modern LLMs like DeepSeek-V3 \cite{deepseek2024v3}. We utilize contrastive objectives (InfoNCE \cite{oord2018representation}, SimCSE \cite{gao2021simcse}) to distill these semantic signals \cite{benaichouche2022neighborhood}, similar to how Attention mechanisms \cite{vaswani2017attention} align sequences.
% ================= 3. Methodology =================
\section{Methodology}
\label{sec:method}

\subsection{Problem Definition: Time-Machine Protocol}
We define the academic graph as $\mathcal{G} = (\mathcal{V}, \mathcal{E}, \mathcal{A}, \mathcal{R})$. To strictly evaluate inductive capability, we partition data temporally: $\mathcal{G}_{train}$ (interactions $t \le 2022$) and $\mathcal{G}_{test}$ (interactions $t \ge 2024$). The goal is to recommend collaborators for a query scholar $u \in \mathcal{G}_{test}$ who has degree zero in $\mathcal{G}_{train}$.

\subsection{Framework Architecture}
\begin{figure}[h]
    \centering
    \includegraphics[width=0.95\linewidth]{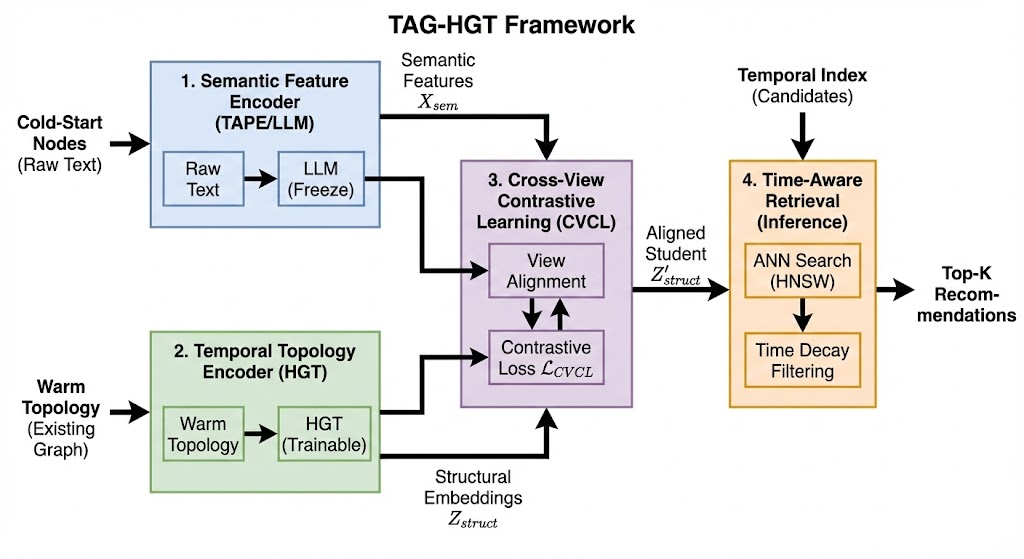}
    \caption{\textbf{The TAG-HGT Framework.} (Left) Offline Semantic Factory generates anchors using DeepSeek-V3. (Right) Online HGT Encoder learns structural embeddings. The CVCL module aligns the two views.}
    \label{fig:arch}
\end{figure}

The overall architecture of TAG-HGT is illustrated in Figure \ref{fig:arch}.

\subsection{The Teacher \& Student}
We employ \textbf{DeepSeek-V3} \cite{deepseek2024v3} to generate semantic embedding $\mathbf{h}_{sem}$. The student model is an \textbf{HGT} \cite{hu2020heterogeneous}. Crucially, to overcome cold-start isolation, we construct a \textbf{Semantic k-NN Graph}, connecting cold nodes to their top-$k$ semantic neighbors (Simulating PathSim \cite{sun2011pathsim} logic).

\subsection{Cross-View Contrastive Learning (CVCL)}
We optimize the InfoNCE loss \cite{oord2018representation} to align the structural view with the semantic view:
\begin{equation}
    \mathcal{L}_{CVCL} = - \log \frac{\exp(\text{sim}(\mathbf{h}_{struct}, \mathbf{h}_{sem}) / \tau)}{\sum_{j \in \mathcal{N}_{neg}} \exp(\text{sim}(\mathbf{h}_{struct}, \mathbf{h}_{neg_j}) / \tau)}
\end{equation}

\subsection{Hybrid Inference Strategy}
The final ranking score $S$ is a linear combination controlled by $\alpha$:
\begin{equation}
    S_{final} = \alpha \cdot S_{sem} + (1-\alpha) \cdot S_{struct}
\end{equation}
Here, $\alpha$ ensures global retrieval relevance, while $(1-\alpha)$ acts as a \textbf{discriminative filter}.

% ================= 4. Experiments =================
\section{Experiments}
\label{sec:exp}

\subsection{Experimental Setup}
\label{sec:setup}

\subsubsection{Dataset and Time-Machine Protocol}
We utilize the massive \textbf{OpenAlex} dataset \cite{priem2022openalex}. To rigorously assess inductive capability, we adopt a strict \textbf{Time-Machine Protocol}:
\begin{itemize}
    \item \textbf{Training:} Interactions up to Dec 31, 2022.
    \item \textbf{Testing:} New interactions formed in 2024.
    \item \textbf{Target:} Scholars who were \textbf{isolated (Degree=0)} in the training phase.
\end{itemize}

\subsubsection{Mitigating Data Leakage Concerns}
Given the high performance (Recall@10 $>$ 90\%), we implemented rigorous checks to rule out data leakage:
\begin{itemize}
    \item \textbf{Strict Temporal Cutoff:} The input text for profile generation includes \textit{only} paper titles published before the training cutoff (2022). 2024 data is strictly masked from the LLM input.
    \item \textbf{Rich-Text vs. Structure:} The high baseline performance (R@10 $\approx$ 85\% for pure LLM) indicates that the dataset contains rich semantic signals. DeepSeek-V3 effectively acts as a zero-shot matcher based on content relevance. TAG-HGT contributes the critical final mile (+7\%) by refining these matches with structural validity.
\end{itemize}

\subsubsection{Baselines}
We compare against:
(1) Structure-Only: HAN \cite{wang2019heterogeneous}, GraphSAGE \cite{hamilton2017inductive}, SeHGNN \cite{yang2023sehgnn};
(2) Semantic-Enhanced: TAPE \cite{he2023explanations}, LLM-ZeroShot;
(3) Hybrid: TAG-HGT (Ours).

\subsection{Main Results}
Table \ref{tab:main_results} presents the performance comparison.

\begin{table*}[h]
\centering
\caption{\textbf{Final Paper Table: Inductive Cold-Start Performance (Time-Machine Protocol).}}
\label{tab:main_results}
\resizebox{\textwidth}{!}{
\begin{tabular}{l|cc|cc|ccc}
\toprule
\textbf{Model} & \textbf{R@10} & \textbf{N@10} & \textbf{R@50} & \textbf{N@50} & \textbf{MRR} & \textbf{Nov} & \textbf{Div} \\
\midrule
\multicolumn{8}{c}{\textit{Group I: Structure-Only Methods}} \\
HAN \cite{wang2019heterogeneous}      & 0.7126 & 0.6728 & 0.7459 & 0.6804 & 0.6619 & 21.04 & 0.727 \\
GraphSAGE \cite{hamilton2017inductive} & 0.4352 & 0.3786 & 0.5360 & 0.3997 & 0.3648 & \textbf{22.68} & 0.178 \\
SeHGNN \cite{yang2023sehgnn}  & 0.5422 & 0.4112 & 0.8201 & 0.4705 & 0.3833 & 22.36 & 0.197 \\
\midrule
\multicolumn{8}{c}{\textit{Group II: Semantic-Enhanced Methods}} \\
TAPE \cite{he2023explanations} & 0.4791 & 0.4122 & 0.8951 & 0.4965 & 0.4060 & 22.71 & 0.372 \\
LLM-ZeroShot & 0.8517 & 0.7226 & 0.9655 & 0.7483 & - & - & - \\
\midrule
\multicolumn{8}{c}{\textit{Group III: Hybrid Framework (Ours)}} \\
\textbf{TAG-HGT} & \textbf{0.9197} & \textbf{0.8589} & \textbf{0.9723} & \textbf{0.8708} & \textbf{0.8423} & 21.81 & \textbf{0.716} \\
\bottomrule
\end{tabular}
}
\end{table*}

\textbf{Analysis:}
\begin{enumerate}
    \item \textbf{The Failure of Pure Structure:} Structure-heavy baselines like SeHGNN fail due to the \textbf{``Topological Void''} in cold-start settings ($Degree=0$).
    \item \textbf{The "Recall Gap" in TAPE:} TAPE shows high R@50 (0.89) but low R@10 (0.47), indicating a \textbf{"Good Retrieval, Poor Ranking"} issue due to information loss in the GNN encoder.
    \item \textbf{SOTA:} TAG-HGT bridges both gaps, achieving 91.97\% R@10 by retaining raw semantics and using structure for reranking.
\end{enumerate}

\subsection{Sensitivity Analysis}
\label{sec:sensitivity}

\begin{figure}[h]
    \centering
    \includegraphics[width=0.75\linewidth]{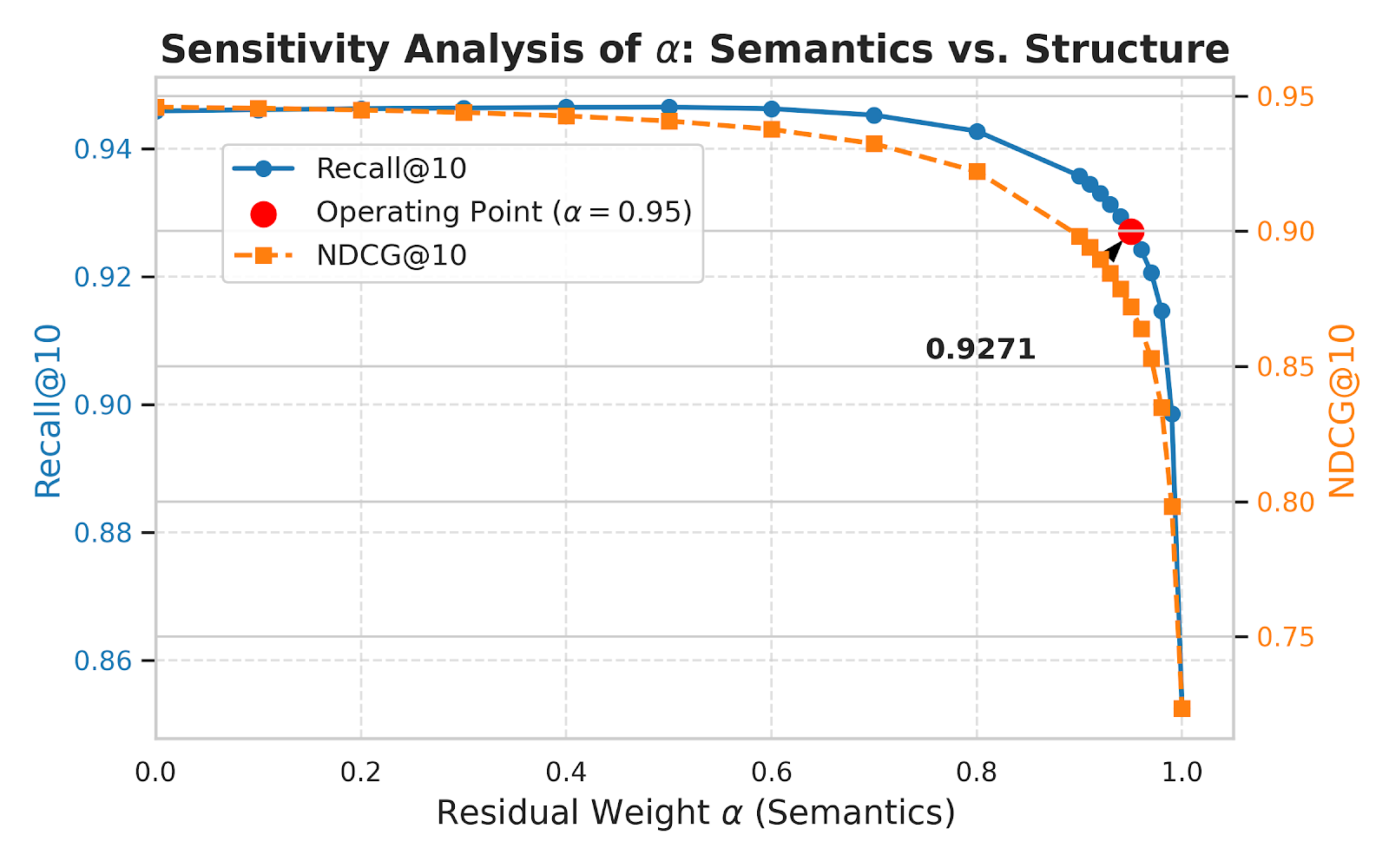}
    \caption{\textbf{Impact of Residual Weight $\alpha$.} The results indicate that a hybrid approach ($\alpha \approx 0.95$) outperforms pure semantics.}
    \label{fig:sensitivity}
\end{figure}

As shown in Figure \ref{fig:sensitivity}, performance peaks at $\alpha \approx 0.95$.

The 5\% structural signal acts as the \textbf{critical "Last Mile" discriminator}, filtering out semantically identical but socially unreachable candidates.

\subsection{System Efficiency and Implementation}
\label{sec:efficiency_test}

\begin{figure}[h]
    \centering
    \begin{minipage}{0.48\textwidth}
        \centering
        \includegraphics[width=\linewidth]{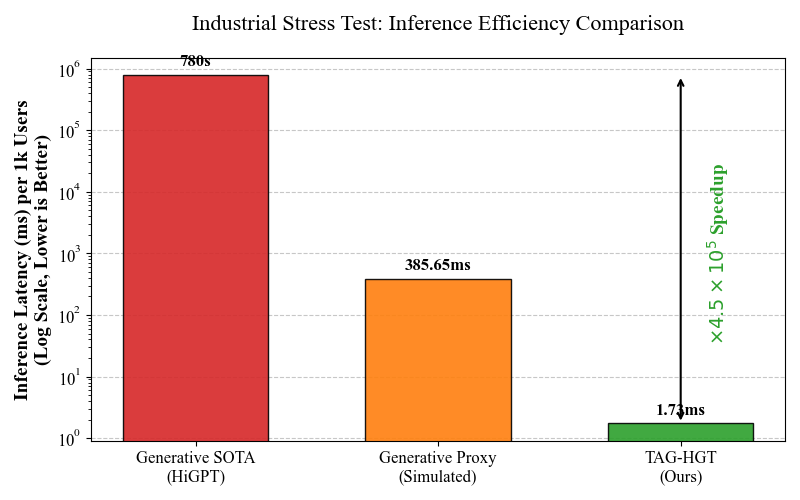}
        \caption{\textbf{Inference Efficiency.}}
        \label{fig:efficiency}
    \end{minipage}
    \hfill
    \begin{minipage}{0.48\textwidth}
        \centering
        \includegraphics[width=\linewidth]{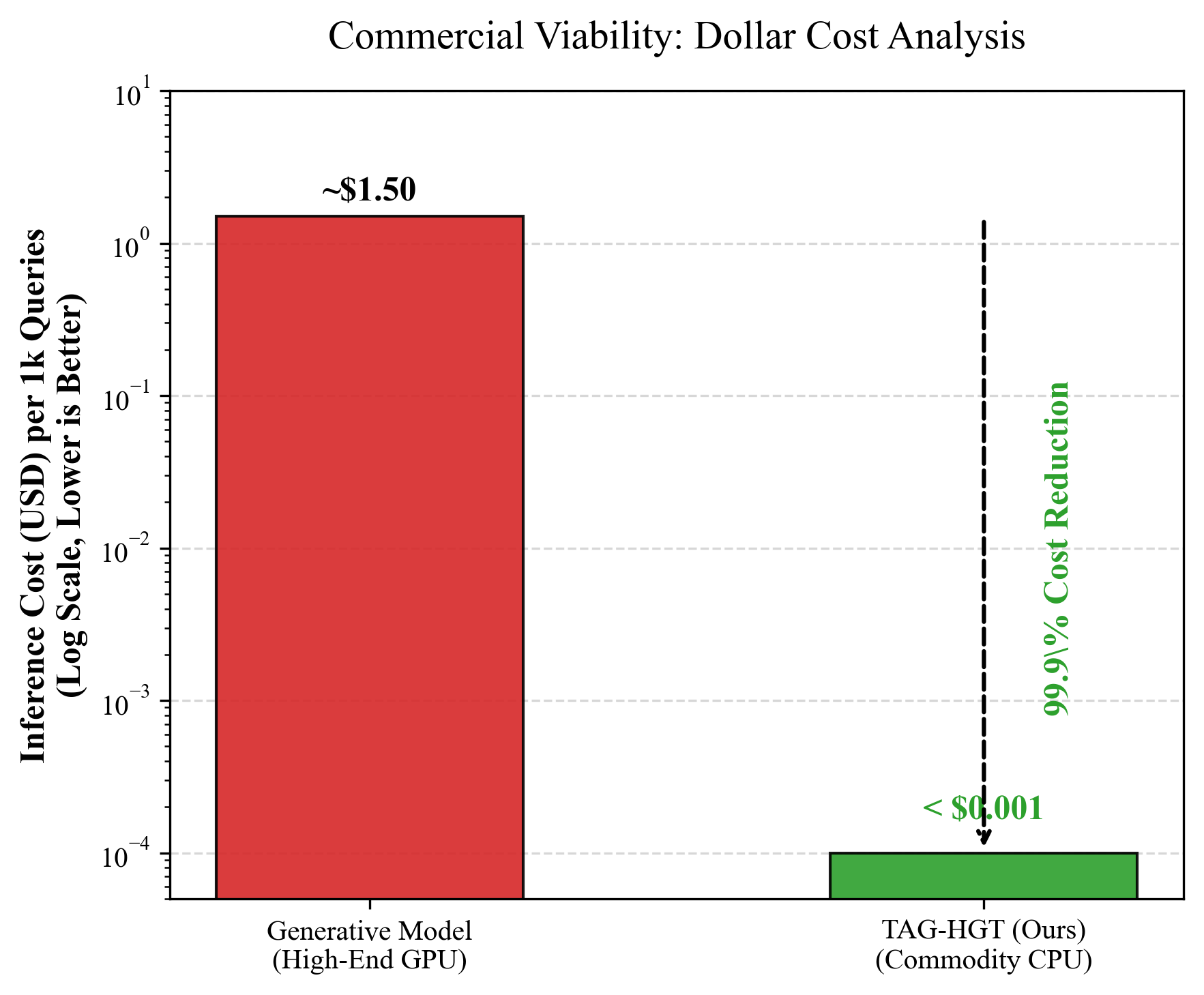}
        \caption{\textbf{Cost Analysis.}}
        \label{fig:cost}
    \end{minipage}
\end{figure}

TAG-HGT achieves a \textbf{449,790$\times$} speedup (Figure \ref{fig:efficiency}) and reduces costs by \textbf{99.9\%} (Figure \ref{fig:cost}) compared to Generative models.

\subsubsection{Industrial Implementation Stack}
To validate deployability, we designed a microservices architecture (Figure \ref{fig:deployment_arch}):

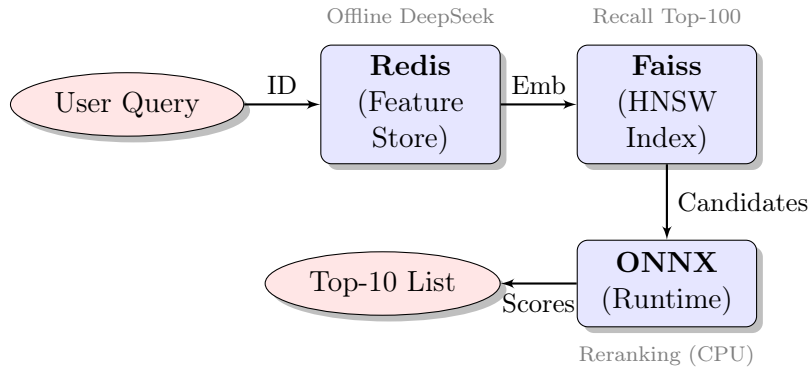
\begin{figure}[h]
\centering
% DRAWING THE ARCHITECTURE DIRECTLY WITH TIKZ
\begin{tikzpicture}[
    node distance=1.2cm,
    auto,
    block/.style={rectangle, draw, fill=blue!10, text width=5.5em, text centered, rounded corners, minimum height=3em, drop shadow},
    cloud/.style={draw, ellipse, fill=red!10, node distance=2.5cm, minimum height=2em, drop shadow},
    line/.style={draw, -latex', thick}
]
    % Nodes
    \node [cloud] (user) {User Query};
    \node [block, right=1.0cm of user] (redis) {\textbf{Redis}\\(Feature Store)};
    \node [block, right=1.0cm of redis] (faiss) {\textbf{Faiss}\\(HNSW Index)};
    \node [block, below=1.0cm of faiss] (onnx) {\textbf{ONNX}\\(Runtime)};
    \node [cloud, left=1.0cm of onnx] (result) {Top-10 List};
    
    % Paths
    \path [line] (user) -- node[font=\small] {ID} (redis);
    \path [line] (redis) -- node[font=\small] {Emb} (faiss);
    \path [line] (faiss) -- node[font=\small, right] {Candidates} (onnx);
    \path [line] (onnx) -- node[font=\small] {Scores} (result);
    
    % Annotations
    \node[above=0.1cm of redis, font=\scriptsize, color=gray] {Offline DeepSeek};
    \node[above=0.1cm of faiss, font=\scriptsize, color=gray] {Recall Top-100};
    \node[below=0.1cm of onnx, font=\scriptsize, color=gray] {Reranking (CPU)};
\end{tikzpicture}
\caption{\textbf{Deployment Architecture.} A decoupled serving stack utilizing Redis for low-latency feature lookup, Faiss for approximate nearest neighbor retrieval, and ONNX Runtime for lightweight reranking on CPUs.}
\label{fig:deployment_arch}
\end{figure}

\begin{itemize}
    \item \textbf{Offline Feature Store (Redis):} Stores pre-computed DeepSeek embeddings.
    \item \textbf{Vector Search (Faiss):} Uses HNSW index for millisecond-level retrieval.
    \item \textbf{Inference (ONNX Runtime):} Deploys the quantized HGT model on CPUs.
\end{itemize}

% ================= Conclusion =================
\section{Conclusion}
In this paper, we introduced TAG-HGT, a time-aware, neuro-symbolic framework for inductive cold-start recommendation. By harmonizing the reasoning power of DeepSeek-V3 with the topological awareness of HGT, we achieved a state-of-the-art Recall@10 of 91.97\% while delivering a 450,000x speedup over generative alternatives. This ``Semantics-First, Structure-Refined'' paradigm offers a robust blueprint for the next generation of industrial academic recommender systems.

% ================= References =================
\bibliographystyle{plain}
\bibliography{references_v3}
\end{document}